\documentclass[
aps,%
12pt,%
final,%
notitlepage,%
oneside,%
onecolumn,%
nobibnotes,%
nofootinbib,%
superscriptaddress,%
noshowpacs,%
centertags]%
{revtex4}

\usepackage{graphicx}
\graphicspath{{./}{Figures/}}
\usepackage{bm}
\usepackage{xcolor}
\usepackage{cmap}
\usepackage{placeins}

\usepackage{amsmath}

\let\oldsection\section
\renewcommand{\section}{
	\renewcommand{\theequation}{\thesection.\arabic{equation}}
	\oldsection}
\let\oldsubsection\subsection
\renewcommand{\subsection}{
	\renewcommand{\theequation}{\thesubsection.\arabic{equation}}
	\oldsubsection}

\begin{document}

\title{Impact of Stochastic Acceleration on the Pickup Proton Distribution and ENA Fluxes in the Heliosphere}

\author{\firstname{I.~I.}~\surname{Baliukin}}
\email{igor.baliukin@cosmos.ru}
\email{igor.baliukin@gmail.com}
\affiliation{Space Research Institute of RAS, Moscow, 117485 Russia}
\affiliation{Lomonosov Moscow State University, Moscow Center for Fundamental and Applied Mathematics, GSP-1, Leninskie Gory, Moscow, 119991 Russia}



\begin{abstract}

The paper describes a kinetic model of pickup protons' distribution in the heliosphere that considers the process of stochastic acceleration by the small-scale Alfvenic and large-scale magnetosonic fluctuations. Using this model, we investigated the impact of this process on the velocity distribution function of pickup protons. The model results showed that stochastic acceleration leads to diffusion in the velocity space and the formation of a high-velocity ``tail'' in the distribution function of pickup protons. We also simulated the energy spectra of energetic neutral atom (ENA) fluxes from the inner heliosheath and compared them to the observations of IBEX-Lo/Hi instruments onboard the IBEX spacecraft in Earth's orbit. The comparison showed that (a) consideration of the large-scale oscillations of the solar wind velocity allows for explaining the IBEX-Hi observations at the highest energy steps, and (b) the velocity diffusion is responsible for an ENA flux increase in the 0.1 -- 0.5 keV range, partially reconciling existing discrepancies between IBEX-Lo observations and models that do not consider this process. 

\vspace{2 ex}
\end{abstract}

\maketitle


\section{INTRODUCTION} \label{sec:introduction}

The supersonic solar wind (SW) interaction with the local interstellar medium (LISM) forms a complex region known as the heliospheric interface. The solar wind is a continuous flow of plasma from the Sun, that expands outwards until it encounters the LISM, forming distinct boundaries. The heliospheric termination shock (TS) is the boundary, where the supersonic SW abruptly slows to subsonic speeds. Another boundary is the heliopause (HP), a tangential discontinuity that separates the SW and LISM plasmas. The region between the TS and the HP is known as the inner heliosheath (IHS). 

In addition to the population of thermal solar wind protons in the heliosphere, there is a suprathermal population of pickup protons that originate from the ionization (due to the charge exchange, photoionization, or electron impact) of interstellar hydrogen atoms, which penetrate the heliosphere from the LISM \citep[the H atoms mean free path for charge exchange is comparable with the characteristic size of the heliosphere;][]{izmod2001}. These newly ionized particles are then ``picked up'' by the SW magnetic field and convected outwards. As they travel through the heliosphere, pickup protons experience stochastic acceleration by interacting with turbulent solar wind environment, namely with small-scale Alfvenic fluctuations and large-scale magnetosonic oscillations of solar wind velocity and magnetic field. The stochastic acceleration significantly modifies the velocity distribution of pickup protons (by generally increasing their energy), and this process was a subject of several studies in the past \citep[see, e.g.][]{fisk1976, isenberg1987, bogdan1991, chalov1995, chalov_fahr1997, fichtner1996, leroux1998}. 

As of today, the global monitoring of plasma properties at the heliospheric boundary is performed only by the Interstellar Boundary Explorer (IBEX) spacecraft \citep{mccomas2009} that orbits the Earth through the measurements of energetic neutral atom (ENA) fluxes using IBEX-Lo \citep[0.01 -- 2 keV;][]{fuselier2009} and IBEX-Hi \citep[0.3 -- 6 keV;][]{funsten2009} instruments. The ENAs originate in the charge exchange process of pickup protons and interstellar hydrogen atoms. These observations carry crucial information on the properties of the region where the solar wind interacts with the interstellar medium. Since the population of pickup protons is parent to the ENAs, a model of pickup protons' distribution in the heliosphere that considers all relevant physics is necessary for correctly interpreting these data. Recently \citet{zirnstein2018a, zirnstein2018b} studied the effect of stochastic acceleration in the inner heliosheath and analyzed the ENA fluxes observed by IBEX in the direction of the heliospheric tail to constrain the proton distribution.

Within this work, we advance the model of pickup protons transport in the heliosphere developed previously by our group \citep{baliukin2020, baliukin2022, baliukin2023} by considering the effect of stochastic acceleration that leads to the velocity diffusion. We emphasize the importance of velocity diffusion in the supersonic solar wind and inner heliosheath, as well as quantify its effect on the velocity distribution of pickup protons and ENA fluxes from the IHS. In particular, we focus on the large (1 -- 2 orders of magnitude) gap between IBEX-Lo observations and predictions of the models in the 0.1 -- 0.5 keV range reported recently for all sky directions \citep{galli2023}. We aim to investigate whether the process of velocity diffusion, which was neglected in the aforementioned models, explains this discrepancy.

Section 2 describes the model and methodology. The results of simulations and comparison with IBEX data are presented in Section 3. Section 4 provides the main conclusions of the work.

\section{MODEL} \label{sec:model}

\subsection{Formulation of the problem}

First, we assume plasma and neutral distributions in the heliosphere are known from the global kinetic-magnetohydrodynamic (kinetic-MHD) model simulations. Within this work, we utilize the distributions obtained using the model developed by S. D. Korolkov, which assumes the same boundary conditions as \citet{izmod2020}. Note that this model uses a singe-fluid approach for the description of all charged particles (thermal protons, pickup protons, electrons, alpha particles, helium ions).

The kinetic equation for the isotropic velocity distribution function of pickup protons $f(t, \mathbf{r}, w)$ in the solar wind plasma reference frame ($\mathbf{v} = \mathbf{V} + \mathbf{w}$), which is also known as Parker transport equation, can be written in the following form:
\begin{equation}
\frac{\partial f}{\partial t} + \mathbf{V} \cdot \frac{\partial f}{\partial \mathbf{r}} = \frac{w}{3} {\rm div}\mathbf{V} \frac{\partial f}{\partial w} 
 + \frac{1}{w^2} \frac{\partial}{\partial w} \left(w^2 D \frac{\partial f}{\partial w} \right)
+ S - f L,
\label{eq:kinetic}
\end{equation}
where $\mathbf{V}$ is the plasma bulk velocity, $w$ is pickup velocity in the solar wind reference frame, and spatial diffusion, which is of minor importance for $\sim$keV particles, is neglected. The first term on the right of this equation is responsible for the change of velocity distribution function due to the adiabatic heating/cooling, which was studied in detail by \citet{baliukin2023}. The second term describes the effect of velocity diffusion -- the process under consideration within this work. In the most general case, the velocity diffusion coefficient $D(t, \mathbf{r}, w)$ can depend on the time, position, and velocity. In equation (\ref{eq:kinetic}), $S(t, \mathbf{r}, w)$ and $L(t, \mathbf{r}, w)$ are source and loss terms due to charge exchange process of solar wind protons with hydrogen atoms that fill the heliosphere (to not overcharge the paper with expressions, we refer to \citet{baliukin2020}, see its equations 4 and 5).

In our simulations, we set boundary conditions at $r_0 = $1 au from the Sun, where we assume that there are no pickup protons, i.e. $f(t, \mathbf{r}_0, w) \equiv 0$. At the heliospheric termination shock, we make use of Liouville's theorem (phase space flow conservation over the shock), the conservation of the magnetic moment (first adiabatic invariant), and the assumption of the weak scattering, as it was done in \citet{baliukin2020}. We utilize the following jump condition \citep{fahr_siewert2011, fahr_siewert2013}:
\begin{equation}
    f^{\mathrm{d}}(t, \mathbf{r}, w) = \frac{s}{C^{3/2}} f^\mathrm{u} \left( t, \mathbf{r}, \frac{w}{C^{1/2}} \right),
    \label{eq:jump_condition}
\end{equation}
where $f^\mathrm{d}$ and $f^\mathrm{u}$ are velocity distribution functions downstream and upstream of the termination shock, $s(t, \mathbf{r}) = n_\mathrm{p}^{\mathrm{d}} / n_\mathrm{p}^{\mathrm{u}}$ is the local shock compression ratio,
\begin{equation}
     C(s, \psi) = \frac{2 A(s, \psi) + B(s, \psi)}{3}, \: A(s, \psi) = \sqrt{\cos^2\psi + s^2 \sin^2\psi},\: B(s, \psi) = \frac{s^2}{A^2},
\end{equation}
and $\psi(t, \mathbf{r})$ is the local upstream shock normal angle. To study the effect of velocity diffusion, we do not include the energetic population of protons accelerated at the TS, which provides a high-energy tail in the velocity distribution, as opposed to what was done in \citet{baliukin2022}.

The number density and temperature of pickup protons are moments of the isotropic velocity distribution function, and they can be calculated as
\begin{align}
    n(t, \mathbf{r}) &= 4\pi \int_0^{+\infty} f(t, \mathbf{r}, w) w^2 \mathrm{d}w, \label{eq:moment_n}\\
    T(t, \mathbf{r}) &= \frac{4\pi m_{\mathrm{p}}}{3 k_{\mathrm{B}} n(t, \mathbf{r})} \int_0^{+\infty}f(t, \mathbf{r}, w) w^4 \mathrm{d}w, \label{eq:moment_T}
\end{align}
where $m_{\mathrm{p}}$ is the proton mass, and $k_{\mathrm{B}}$ is the Boltzmann constant. The pressure of pickup protons can be calculated as $p(t, \mathbf{r}) = n(t, \mathbf{r}) k_{\mathbf{B}} T(t, \mathbf{r})$. Therefore, it can be shown that usage of jump condition (\ref{eq:jump_condition}) leads to
\begin{equation}
    \frac{n^{\mathrm{d}}}{n^{\mathrm{u}}} = s,\:
    \frac{T^{\mathrm{d}}}{T^{\mathrm{u}}} = C, \:
    \frac{p^{\mathrm{d}}}{p^{\mathrm{u}}} = s C.
\end{equation}

In the velocity space, we utilize the following boundary conditions:
\begin{equation}
    \frac{\partial f}{\partial w} \bigg|_{w = 0} =  \frac{\partial f}{\partial w} \bigg|_{w \rightarrow +\infty} = 0.
    \label{eq:boundary}
\end{equation}
The first condition comes from the assumption that the velocity distribution function is isotropic in the plasma reference frame, and for the high velocities we assume a soft boundary condition.

In this work, we consider the stationary problem, so the variable $t$ and its derivatives are omitted hereafter. Since $\mathbf{V} = V(s) \mathbf{e}_s$, the convective term on the left of equation (\ref{eq:kinetic}) can be rewritten as 
\begin{equation}
    \mathbf{V} \cdot \frac{\partial f}{\partial \mathbf{r}} = V \frac{\partial f}{\partial s} = \frac{\partial f}{\partial \tau},
\end{equation}
where $s$ is a coordinate along the streamline ($s = 0$ at the Sun) and $\tau$ is convective time defined as
\begin{equation}
    \tau(s) = \int_{s_0}^s \frac{\mathrm{d}s}{V(s)}.
\end{equation}
Therefore, the kinetic equation (\ref{eq:kinetic}) takes the form
\begin{equation}
    \frac{\partial f}{\partial \tau} 
    = \frac{w}{3} {\rm div}\mathbf{V} \frac{\partial f}{\partial w} 
    + \frac{1}{w^2} \frac{\partial}{\partial w} \left(w^2 D \frac{\partial f}{\partial w} \right)
    + S - f L,
\label{eq:kinetic2}
\end{equation}
where $f = f(\tau, w)$, $D = D(\tau, w)$, $S = S(\tau, w)$, and $L = L(\tau, w)$.

\subsection{Diffusion coefficients}

\subsubsection{Supersonic solar wind region}

In our simulations, the diffusion coefficients in the supersonic SW region are taken in accordance with the works of \citet{chalov1997, chalov2003, chalov2006}. We consider the combined effect of small-scale Alfvenic and large-scale magnetosonic fluctuations on the formation of pickup proton energy spectrum, i.e., $D = D_{\mathrm{A}} + D_{\mathrm{m}}$:
\begin{itemize}
    \item The Alfvenic turbulence can be generated in the distant solar wind by different instabilities, such as the instability of the ring velocity distribution of freshly originated pickup ions. The velocity diffusion coefficient $D_{\mathrm{A}}$ can be derived from the quasi-linear theory of the cyclotron resonant wave-particle interaction \citep{isenberg1987, chalov1997}:
    \begin{equation}
        D_{\mathrm{A}}(\mathbf{r}, w) = \frac{\pi^2 (q-1)}{q(q+2)} \left( \frac{\Omega L_{\mathrm{A}}}{2 \pi w} \right)^{2-q} \frac{v^2_{\mathrm{A}} w}{L_{\mathrm{A}}} \frac{\langle \delta B^2 \rangle}{B^2},
        \label{eq:diffusion_alfvenic}
    \end{equation}
    where $q$ is the spectral index of turbulence ($5/3$ is adopted in our work), $\Omega = e B / m_{\mathrm{p}}$ is the proton cyclotron frequency, $e$ is the proton charge, $B$ is the magnitude of the interplanetary magnetic field, $v_{\mathrm{A}}$ is the Alfven velocity, $L_{\mathrm{A}}$ is the correlation length of Alfvenic turbulence, and $\langle \delta B^2 \rangle$ is the mean-square amplitude of magnetic field fluctuations. We assume that
    \begin{equation}
        \langle \delta B^2 \rangle = \langle \delta B^2 \rangle_{\mathrm{E}} \left( \frac{r_{\mathrm{E}}}{r} \right)^\alpha,\: L_{\mathrm{A}} = L_{\mathrm{AE}} \left( \frac{r}{r_{\mathrm{E}}} \right)^\delta,
    \end{equation}
    where $\alpha = 3$ \citep{hollweg1974}, $\delta = 1$ \citep{jokipii1973}, $L_{\mathrm{AE}} = 0.01$ au \citep{jokipii_coleman_1968}, and $r_{\mathrm{E}} = 1$ au. We also introduce the parameter $\zeta_{\mathrm{AE}} = \langle \delta B^2 \rangle_{\mathrm{E}} / B^2_{\mathrm{E}}$, where $\langle \delta B^2 \rangle_{\mathrm{E}}$ and $B_{\mathrm{E}}$ are the mean-square amplitude of magnetic field fluctuations and the magnetic field magnitude at the Earth's orbit. 

    \item The large-scale oscillations in the magnitudes of solar wind velocity are usually connected with corotating and merged interaction regions and contain the structures of large-scale interplanetary shock waves. The acceleration of particles by this type of fluctuation is equivalent to the second-order Fermi acceleration. The velocity diffusion coefficient $D_{\mathrm{m}}$ can be calculated as
    \begin{equation}
        D_{\mathrm{m}}(\mathbf{r}, w) = \frac{\langle \delta V^2 \rangle^{1/2} w^2}{9 L_{\mathrm{m}}},
        \label{eq:diffusion_magnetosonic}
    \end{equation}
    where $\langle \delta V^2 \rangle$ is the mean-square amplitude of solar wind velocity and $L_{\mathrm{m}}$ is the correlation length of large-scale magnetosonic fluctuations. We assume that
    \begin{equation}
        \langle \delta V^2 \rangle = \langle \delta V^2 \rangle_{\mathrm{E}}  \left( \frac{r_{\mathrm{E}}}{r} \right)^{2 \beta},
    \end{equation}
    $\beta = 0.7$ and $L_{\mathrm{m}} = $ 3 au \citep[based on the observations by Pioneer and Voyager spacecraft;][]{whang1988, whang1990, richardson2001}. We also introduce the parameter $\zeta_{\mathrm{mE}} = \langle \delta V^2 \rangle_{\mathrm{E}}^{1/2} / V_{\mathrm{E}}$, where $V_{\mathrm{E}}$ and $\langle \delta V^2 \rangle_{\mathrm{E}}$ are the solar wind velocity and mean-square amplitude of its oscillations at the Earth's orbit. 
    
\end{itemize}

We note that $D_{\mathrm{A}} \sim w^{2/3}$ and $D_{\mathrm{m}} \sim w^2$, as can be seen from equations (\ref{eq:diffusion_alfvenic}) and (\ref{eq:diffusion_magnetosonic}). Therefore, the small-scale Alfvenic turbulence is responsible for the diffusion of low-energy particles, while the large-scale magnetosonic oscillations are more effective for higher energies. We also note that in the model of diffusion coefficients described above there are two free parameters, $\zeta_{\mathrm{AE}}$ and  $\zeta_{\mathrm{mE}}$, which characterize the magnitudes of the Alfvenic and magnetosonic fluctuations at the Earth's orbit.

\subsubsection{Inner heliosheath region}

Due to the lack of information about the spatial behavior and properties of Alfvenic or magnetosonic fluctuations in the inner heliosheath (IHS), in this region, we use the empirical diffusion coefficient estimated based on the comparison with the IBEX data. The dependence of the diffusion coefficient on the proton velocity $w$ in the plasma reference frame was taken in the same power-law form as in \citet{zirnstein2018a, zirnstein2018b}, namely:
\begin{equation}
    D(w) = D_{\mathrm{IHS,0}} \left( \frac{w}{w_0} \right)^\gamma, 
    \label{eq:diffusion_ihs}
\end{equation}
where $w_0$ = 1 km/s. \citet{zirnstein2018a} performed the parametric fitting and found that the spectral index $\gamma$ = 1.25 and $D_{\mathrm{IHS,0}} = D_{\mathrm{Z}} = 1.1 \times 10^{-2}$ m$^2$/s$^3$ allows explaining the IBEX data the best. We use these values as a reference in our work, even though we note that they were obtained specifically for the direction of the heliospheric tail.

We assume in our calculations that the dependence (\ref{eq:diffusion_ihs}) is true everywhere in the IHS. The numerical simulations were performed in three cases:
\begin{enumerate}
    \item $D_{\mathrm{IHS,0}} = 0$ (no velocity diffusion); 
    \item $D_{\mathrm{IHS,0}} = D_{\mathrm{Z}}$ (moderate velocity diffusion);
    \item $D_{\mathrm{IHS,0}} = 10 \times D_{\mathrm{Z}}$ (strong velocity diffusion),
\end{enumerate}
and the spectral index $\gamma$ was not varied. 

\subsection{Numerical method}

At the first step, we reconstruct the streamline by solving the equation of motion $\mathrm{d}\mathbf{r}/\mathrm{d}t = \mathbf{V}$ numerically backward in time from a point of interest $\mathbf{r}$ in the heliosphere to 1 au from the Sun, where $s = s_0 = 1$ au ($\tau = 0$) and boundary condition is set. Then, the kinetic equation (\ref{eq:kinetic2}) is solved numerically using the operator splitting approach, also known as the method of fractional steps \citep{yanenko1967}. We note that the other commonly used method of solving the Fokker-Planck type equation (\ref{eq:kinetic2}) is based on transforming the partial differential equation into stochastic differential equations \citep[see, e.g.][]{chalov1997}.

The right side of equation (\ref{eq:kinetic2}) can be written as the sum of two operators:
\begin{align}
    \frac{\partial f}{\partial \tau} &= \mathcal{L}_1(f) + \mathcal{L}_2(f),\\
    \mathcal{L}_1(f) &= \frac{w}{3} {\rm div}\mathbf{V} \frac{\partial f}{\partial w} + S - f L, \\
    \mathcal{L}_2(f) &= \frac{1}{w^2} \frac{\partial}{\partial w} \left(w^2 D \frac{\partial f}{\partial w} \right),
\end{align}
where operator $\mathcal{L}_1$ accounts for convection in the velocity space and sources/losses, while operator $\mathcal{L}_2$ is responsible for diffusion. To update the values of the function $f$ from timestep $i$ to timestep $i+1$ we do two fractional updatings:
\begin{align}
    f^{i+1/2} &= \mathcal{L}_1(f^i, \Delta \tau),\label{eq:fracstep1}\\ 
    f^{i+1} &= \mathcal{L}_2(f^{i+1/2}, \Delta \tau)\label{eq:fracstep2},
\end{align}
where $i$ is the time (or spatial) index ($\tau^{i} = i \Delta \tau$).

To do the first fractional step (\ref{eq:fracstep1}), we utilize the method of characteristics, which is described in detail in a series of our previous works \citep{baliukin2020, baliukin2022, baliukin2023}. We note that this method allows for avoiding numerical diffusion. For the second fractional step (\ref{eq:fracstep2}), we use the well-proven Crank -- Nicolson finite difference scheme \citep[see, e.g.][]{press2007}, which is unconditionally stable and second-order accurate both in time $\tau$ and velocity $w$:
\begin{align}
    \frac{f_{j}^{i+1} - f_{j}^{i}}{\Delta \tau} = \frac{1}{w_j^2\Delta w} \Biggl[
& w_{j+1/2}^2 D_{j+1/2}^{i+1/2} \left( \frac{1}{2} \frac{f_{j+1}^{i+1} - f_{j}^{i+1}}{\Delta w} + \frac{1}{2} \frac{f_{j+1}^{i} - f_{j}^{i}}{\Delta w} \right) \nonumber\\
-&w_{j-1/2}^2 D_{j-1/2}^{i+1/2} \left( \frac{1}{2} \frac{f_{j}^{i+1} - f_{j-1}^{i+1}}{\Delta w} + \frac{1}{2} \frac{f_{j}^{i} - f_{j-1}^{i}}{\Delta w} \right) \Biggr],
\label{eq:finite}
\end{align}
where $j$ is the velocity index ($j = 0, \dots, N$), $w_{j} = j \Delta w$, $\Delta w = w_{\mathrm{max}} / N$, and
\begin{align}
    & w_{j+1/2} = \frac{w_{j+1} + w_{j}}{2},\: D_{j+1/2}^{i+1/2} = \frac{D(\tau^{i+1}, w_{j+1/2}) + D(\tau^{i}, w_{j+1/2})}{2}, \\
    & w_{j-1/2} = \frac{w_{j} + w_{j-1}}{2},\: D_{j-1/2}^{i+1/2} = \frac{D(\tau^{i+1}, w_{j-1/2}) + D(\tau^{i}, w_{j-1/2})}{2}.
\end{align}

The equation (\ref{eq:finite}) can be rewritten in the form
\begin{equation}
a_{j} f_{j-1}^{i+1} + b_{j} f_{j}^{i+1} + c_{j} f_{j+1}^{i+1} = d_{j},\: j = 1, \dots, N-1,
\label{eq:tridiagonal}
\end{equation}
where the coefficients (for $j = 2, \dots, N-2$) are:
\begin{align}
a_{j} & = -\alpha w_{j-1/2}^2 D_{j-1/2}^{i+1/2}, \nonumber\\
b_{j} & = +\alpha (w_{j+1/2}^2 D_{j+1/2}^{i+1/2} + w_{j-1/2}^2 D_{j-1/2}^{i+1/2}) + 1, \nonumber \\
c_{j} & = -\alpha w_{j+1/2} D_{j+1/2}^{i+1/2}, \nonumber \\
d_{j} & = +\alpha \left(w_{j+1/2}^2 D_{j+1/2}^{i+1/2} (f_{j+1}^i - f_j^i) - w_{j-1/2}^2 D_{j-1/2}^{i+1/2} (f_{j}^i - f_{j-1}^i) \right) + f_{j}^{i},
\end{align}
where $\alpha = \Delta \tau / (2 w_j^2 (\Delta w)^2)$. The coefficients for $j = 0$ and $j = N - 1$ can be obtained similarly using equalities $f_0 = f_1$ and $f_{N} = f_{N-1}$, which follow from the boundary conditions (\ref{eq:boundary}). The system of linear equations (\ref{eq:tridiagonal}) is solved using a tridiagonal matrix algorithm.

We have developed a numerical code to make the second fractional step using the Crank -- Nicolson method. The results of simulations were compared with the analytical solution: the diffusion equation
\begin{equation}
\frac{\partial f}{\partial \tau} = \frac{1}{w^2} \frac{\partial}{\partial w} \left( w^2 D \frac{\partial f}{\partial w} \right) = \mathcal{L}_2(f)
\end{equation}
with initial condition $f(0, w) = f_0 w_0 \delta(w - w_0)$ and $D = \mathrm{const}$ has the following solution \citep[see, e.g.][]{chalov2006}:
\begin{equation}
f(\tau, w) = \frac{f_0 w_0^2}{\sqrt{4 \pi D \tau} w} \left[ \exp \left( -\frac{(w - w_0)^2}{4 D t} \right) - \exp \left( -\frac{(w + w_0)^2}{4 D t} \right) \right].
\end{equation}

\section{RESULTS} \label{sec:results}

In our simulations, we assume that $\Delta s$ is constant, so $\Delta \tau$ changes along the streamline inversely proportional to the plasma velocity $V(s)$. Hereafter we provide the results of calculations with parameters $\Delta s = 10^{-3}$ au, $w_{\mathrm{max}} = 10^4$ km/s, and $N = 5000$.

We performed a set of simulations using different values of $\zeta_{\mathrm{AE}}$, $\zeta_{\mathrm{mE}}$, and $D_{\mathrm{IHS,0}}$ (listed in Table \ref{tab:models}). The parameters ($\zeta_{\mathrm{AE}}$, $\zeta_{\mathrm{mE}}$) characterize the level of velocity diffusion in the supersonic solar wind region, while the parameter $D_{\mathrm{IHS,0}}$ is responsible for diffusion in the inner heliosheath. 

\begin{table}
    \centering
    \caption{The values of parameters ($\zeta_{\mathrm{AE}}$, $\zeta_{\mathrm{mE}}$, $D_{\mathrm{IHS,0}}$), which define the velocity diffusion coefficients, used in different model simulations. SSW = supersonic solar wind, IHS = inner heliosheath.}
    \label{tab:models}
    \begin{tabular}{ccccl} 
        \hline
        Model & $\zeta_{\rm AE}$ & $\zeta_{\rm mE}$ & $D_{\mathrm{IHS,0}}$ & Description\\
        \hline
        0 & 0.0 & 0.0 & 0.0 & No velocity diffusion \\
        \hline
        1 & 0.1 & 0.0 & 0.0 & Only Alfvenic (small-scale) fluctuations in the SSW \\
        2 & 0.0 & 0.5 & 0.0 & Only magnetosonic (large-scale) oscillations in the SSW  \\
        3 & 0.1 & 0.5 & 0.0 & All fluctuations (Alfvenic + magnetosonic) in the SSW \\
        \hline
        4 & 0.0 & 0.0 & $D_{\rm Z}$ & No velocity diffusion in the SSW, moderate diffusion in the IHS \\
        5 & 0.1 & 0.5 & $D_{\rm Z}$ & All fluctuations in the SSW, moderate diffusion in the IHS \\
        6 & 0.1 & 0.5 & $10 D_{\rm Z}$ & All fluctuations in the SSW, strong diffusion in the IHS \\
        \hline
    \end{tabular}
\end{table}

\subsection{Velocity diffusion effect on the distribution function}

\begin{figure}
    \center{\includegraphics[width=1.0\linewidth]{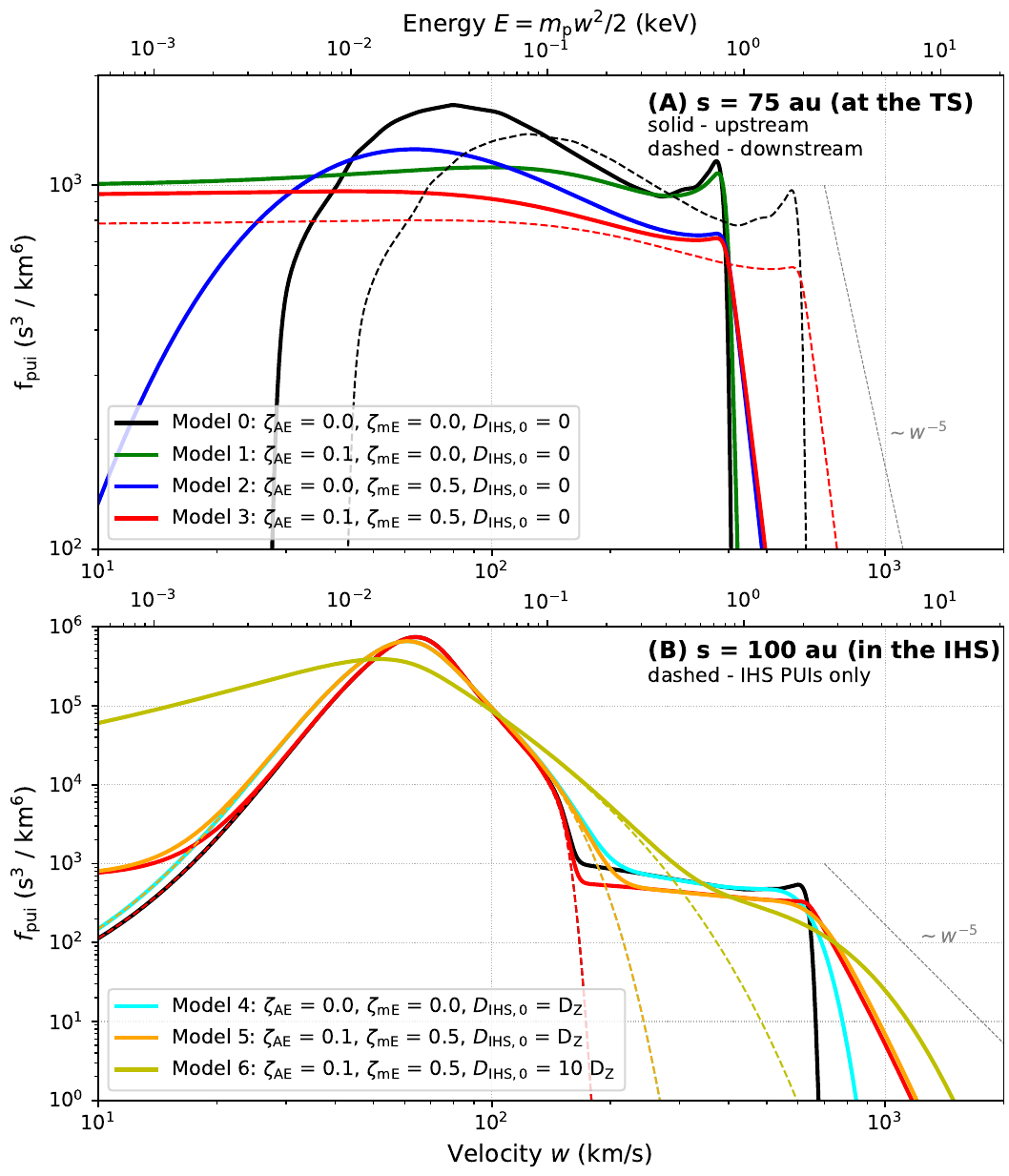}}
    \caption{The velocity distribution function of pickup protons in the solar wind reference frame along the streamline emerging from the Sun in the upwind direction. Panel (A) shows the velocity distribution function at the TS ($s = 75$ au; solid and dashed lines -- upstream and downstream of the TS, respectively), while Panel (B) shows the results of simulations at $s = 100$ au (in the IHS; dashed lines represent the velocity distribution of pickup protons that originated in the inner heliosheath). The black, green, blue, red, cyan, orange, and yellow lines represent the results of simulations of Models 0 -- 6, respectively, with different parameters ($\zeta_{\mathrm{AE}}$, $\zeta_{\mathrm{mE}}$, $D_{\mathrm{IHS,0}}$) compiled in Table \ref{tab:models}. }
    \label{fig:fpui}
\end{figure}

Figure \ref{fig:fpui} shows the velocity distribution function of pickup protons in the solar wind reference frame along the streamline emerging from the Sun in the direction opposite to the local interstellar medium flow (upwind direction). Panel (A) shows the velocity distribution function at the TS ($s = 75$ au in our simulations), while Panel (B) shows the results of simulations at $s = 100$ au (in the IHS). 

As can be seen in Figure \ref{fig:fpui}(A), the Alfvenic small-scale fluctuations are responsible for the velocity diffusion with small velocities (see the black and green lines), while the large-scale magnetosonic fluctuations are more effective in the diffusion of particles with high energies (compare the black and blue lines). The latter is almost solely responsible for the high-velocity tail production (compare the blue and red lines). The effect of compression at the TS, described in our model by the jump condition (\ref{eq:jump_condition}), is seen from the comparison of solid and dashed lines in panel (A) of Figure \ref{fig:fpui}.

In the IHS, the solar wind plasma is decelerated, so the relative velocity between the solar wind protons and interstellar hydrogen atoms is smaller (compared to the supersonic solar wind region). So, in this region, the charge exchange results in the production of pickup protons with $\lesssim$ 0.1 keV energies. The dashed lines in Figure \ref{fig:fpui}(B) show the velocity distribution functions of pickup protons that originated in the IHS. 

The comparison of black and cyan lines in panel (B) of Figure \ref{fig:fpui} shows the effect of the velocity diffusion in the IHS only -- it smooths the sharp cut-off at $\sim$2 keV and provides a high-velocity tail. The other noticeable effect is that diffusion makes the velocity distribution function of IHS pickup protons broader, which results in higher values of velocity distribution function with energies $\sim$0.2 keV. The orange line shows the effect of the velocity diffusion in the IHS added to the diffusion in the supersonic solar wind. The yellow line represents the model simulation with (ten times) increased level of velocity diffusion in the IHS. As can be seen from the comparison of the red, orange, and yellow lines, the diffusion in the velocity space is responsible for the significant increase of pickup protons with kinetic energies $\sim$0.1 -- 1 keV in the solar wind reference frame.

\subsection{Comparison with IBEX ENA flux energy spectra}

\begin{figure}
    \center{\includegraphics[width=1.0\linewidth]{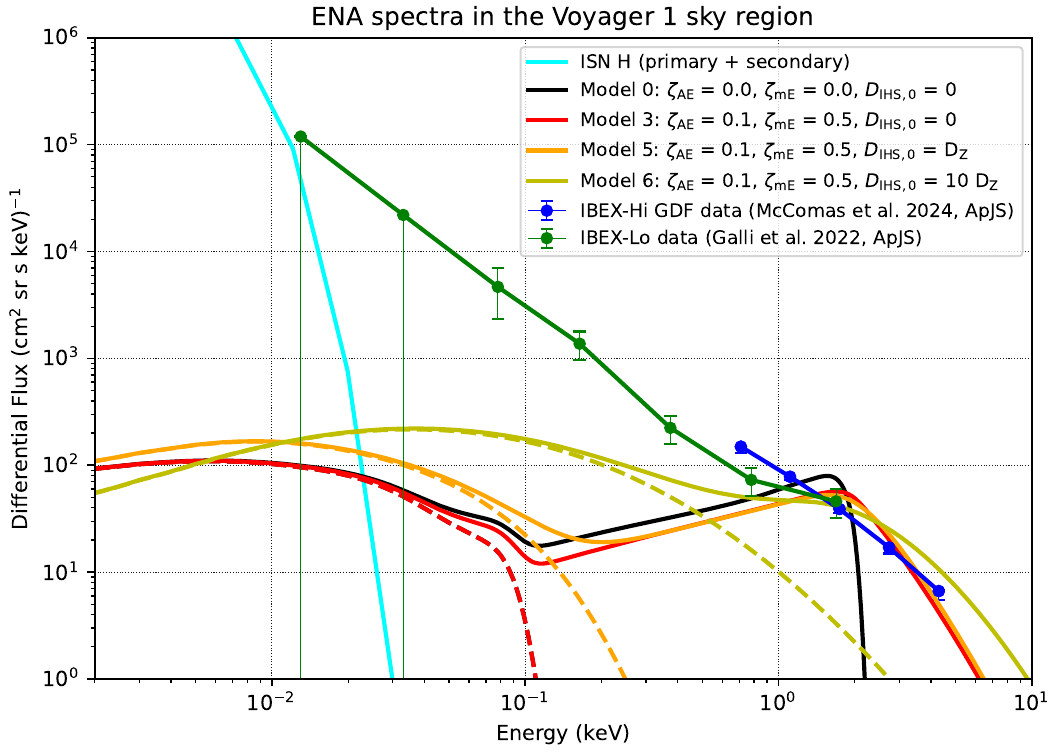}}
    \caption{Energy spectra of ENA hydrogen fluxes in the Voyager 1 region of the sky (see text for the description) as seen by the observer at 1 au. The black, red, orange, and yellow lines represent the results of simulations of Models 0, 3, 5, and 6, respectively, with different parameters ($\zeta_{\mathrm{AE}}$, $\zeta_{\mathrm{mE}}$, $D_{\mathrm{IHS,0}}$) compiled in Table \ref{tab:models} that describe the level of fluctuations. The cyan line shows the fluxes of primary and secondary interstellar hydrogen atoms. The blue lines with points show the IBEX-Hi GDF separated data \citep{mccomas2024}, while the green lines with points show the IBEX-Lo data \citep{galli2022b}. The data are averaged over 11 years (2009 -- 2019).}
    \label{fig:spectra}
\end{figure}

Figure \ref{fig:spectra} shows the energy spectra of ENA hydrogen differential fluxes in the direction of the Voyager 1 region of the sky seen by the observer at 1 au. The blue lines with points show the IBEX-Hi globally distributed flux (GDF) separated data \citep{mccomas2024}, while the green lines with points show the IBEX-Lo data \citep{galli2022b}. The data are averaged over 11 years (2009 -- 2019) to compare with the results of stationary model simulations. 

The black, red, orange, and yellow lines in Figure \ref{fig:spectra} show the ENA fluxes from the inner heliosheath calculated within Models 0, 3, 5, and 6, respectively (see Table \ref{tab:models} for the model description). The fluxes are calculated in the middle of IBEX $6^\circ \times 6^\circ$ pixels and averaged over the region of the sky with longitudes 246$^\circ$–270$^\circ$ and latitudes 30$^\circ$–54$^\circ$ (16 pixels). The cyan line shows the fluxes of primary and secondary interstellar hydrogen atoms that originate outside the heliopause.

As can be seen from the comparison of black and red lines in Figure \ref{fig:spectra}, which differ only by the inclusion of the velocity diffusion in the supersonic solar wind region, the model with this process taken into account provides a better fit to the IBEX-Hi data at the highest energy steps. The velocity diffusion flattens the spectrum that inherited its filled-shell profile with a sharp cutoff from the velocity distribution function of pickup protons (see Figure \ref{fig:fpui}).

Figure \ref{fig:spectra} also shows a big ``gap'' (1 -- 2 orders of magnitude) between the IBEX-Lo data (green line) and model simulations without the velocity diffusion (black line, Model 0) for the energy range 0.1 -- 0.5 keV. This discrepancy was reported recently by \citet{galli2023} not only for Voyager 1 but also for other sky regions (Voyager 2, North and South poles, Port Tail Lobe, and Downwind). We see from the results of model simulations that the increasing velocity diffusion in the inner heliosheath enhances the ENA fluxes in the range of their deficiency (see red, orange, and yellow lines). However, even a strong velocity diffusion (yellow line, Model 6) does not allow closing the gap completely. Moreover, it leads to unrealistically high values of fluxes at the highest energy steps of IBEX-Hi, making the model inconsistent with the data.

We note another source of protons in the IHS -- the population of thermal solar wind protons. However, it has a relatively low temperature in the IHS \citep[$\sim$60 000 K according to Voyager 2 observations; see, e.g.][]{richardson2022} and the charge exchange with these protons do not provide enough atoms with energies 0.1 -- 0.5 keV \citep{galli2023}.

According to our simulations with velocity diffusion taken into account, the population of pickup protons originating in the IHS (dashed lines in Figure \ref{fig:spectra}) is the main contributor to the fluxes in the energy range 0.1 -- 0.5 keV. Nevertheless, their abundance should be higher, and/or their velocity distribution in the solar wind reference frame should be shifted to higher velocities to fill the gap. The latter is controlled mainly by the solar wind velocity in the IHS, which might be different in the models compared to the reality.

The other possibility is that the contribution of ENAs from outside the heliopause may solve the deficiency problem. However, this scenario can be ruled out since (a) the gap between the models and data is present in all sky directions (even from the heliospheric tail), and (b) observations of the IBEX Ribbon \citep{mccomas2009_science} suggest strong anisotropy of the velocity distribution function of pickup protons outside the heliopause.

\section*{CONCLUSIONS}

In this work, we have developed the kinetic model of pickup protons' distribution in the heliosphere that considers the velocity diffusion and studied its effect on the velocity distribution function. We have also simulated the energy spectra of ENA fluxes observed by IBEX-Lo/Hi at 1 au from the Sun, with and without the velocity diffusion process taken into account in the supersonic SW and IHS regions, and compared it to the data. The main conclusions of the work can be summarized as follows.

\begin{enumerate}

    \item The velocity diffusion is responsible for the formation of a ``tail'' in the velocity distribution function of pickup protons in the supersonic SW region. Thus, when the pickup protons reach the heliospheric termination shock, their distribution already has a pronounced ``tail'', which is necessary for pickup protons to enter the regime of the drift acceleration.

    \item The model with large-scale fluctuations in the supersonic SW region taken into account allows explaining the IBEX-Hi data at the highest energy steps, even without the consideration of the energetic population of pickup protons accelerated at the TS.

    \item The velocity diffusion increases the number of pickup protons in the IHS with velocities $\sim$100 -- 300 km/s and leads to an increase in the ENA fluxes in the 0.1 -- 0.5 keV range. Thus, the effect of velocity diffusion can partially explain the difference between the ENA fluxes observed by the IBEX spacecraft and two independent models (Moscow and Boston models) that do not take this process into account \citep{galli2023}.

    \item The differences between the observed and calculated ENA fluxes remain significant even when strong velocity diffusion in the IHS is assumed, suggesting further investigation is needed.

\end{enumerate}

\section*{ACKNOWLEDGEMENTS}	

IB acknowledges the support from the Chinese Academy of Sciences (CAS) President's International Fellowship Initiative (PIFI). The author thanks S.D. Korolkov for providing the global distributions of plasma and hydrogen atoms in the heliosphere and V.V. Izmodenov for fruitful discussions. This work also benefited from discussions and results enabled by the SHIELD DRIVE center (\url{https://shielddrivecenter.com/}).

\bibliographystyle{unsrtnat}
\bibliography{bibliography}

\end{document}